\documentclass[
reprint, amsmath,amssymb,aps,twocolumn]{revtex4-2}
\usepackage[force]{feynmp-auto}
\usepackage{subfigure}
\usepackage{graphicx}
\usepackage{float}
\usepackage{dcolumn}
\usepackage{bm}
\usepackage{physics}
\usepackage[backref]{hyperref}

\begin{document}
\preprint{APS/123-QED}
\title{Phonon Stability of Quantum Droplets in dipolar Bose gases}
\author{Fan Zhang}
\author{Lan Yin}
\email{yinlan@pku.edu.cn}
\affiliation{School of Physics, Peking University, Beijing 100871, China}
\date{\today}
\begin{abstract}
Stabilized by quantum fluctuations, dipolar Bose-Einstein condensates can form self-bound liquid-like droplets.  However in the Bogoliubov theory, there are imaginary phonon energies  in the long-wavelength limit, implying dynamical instability of this system.  A similar instability appears in the Bogoliubov theory of a binary quantum droplet, and is removed due to higher-order quantum fluctuations as shown recently \cite{gu2020phonon}.  In this work, we study the excitation energy of a dipolar quantum droplet in the Beliaev formalism, and find that quantum fluctuations significantly enhance the phonon stability.  We adopt a self-consistent approach without the problem of complex excitation energy in the Bogoliubov theory, and obtain a stable anisotropic sound velocity which is consistent with the superfluid hydrodynamic theory, but slightly different from the result of the extended Gross-Pitaevski (GP) equation due to quantum depletion.  A modified GP-equation in agreement with the Beliaev theory is proposed, which takes the effet of quantum fluctuations into account more completely.
\end{abstract}
\maketitle
Quantum droplets have been realized in different systems, including single-component dipolar Bose gases of $^{164}\rm{Dy}$\cite{Kadau2016b,Ferrier-Barbut2016,Ferrier-Barbut2016a,Schmitt2016a,Wenzel2017},$ ^{166}\rm{Er}$\cite{Chomaz2016}atoms, nonmagnetic binary $ ^{39}\rm{K}$ mixture \cite{Cabrera2018,Cheiney2018,Semeghini2018}, and heteronuclear  $ ^{39}\rm{K}$-$ ^{87}\rm{Rb}$ mixtures\cite{DErrico2019}.  There are also proposals to create quantum droplets in binary dipolar mixtures\cite{bisset2021quantum,smith2021quantum}.  Usually an ultracold bosonic gas is a dilute weakly-interacting system in which the contribution from quantum fluctuations are typically small compared with the mean-field contribution, but quantum droplets are just exceptions. Due to the repulsive Lee-Huang-Yang (LHY) energy \cite{Lee1957} which overcomes the net attractive mean-field energy near the mean-field unstable point, a nonmagnetic binary bosonic mixture displays a self-bound liquid-like state \cite{Petrov2015}, as observed in experiments \cite{Cabrera2018,Cheiney2018,Semeghini2018,DErrico2019}.  The LHY energy \cite{Lima2011} also stabilizes the single-component dipolar quantum droplet in the dipole-dominated regime where the dipolar interaction dominates over the repulsive $s$-wave interaction.  Phase-coherent droplet arrays have also been observed in experiments \cite{Bombin2017,Tanzi2019,Roccuzzo2019,Bottcher2019a}, where broken translational symmetry and superfluid order appear at the same time, consistent with a supersolid \cite {Li2017,Leonard2017}.

In a nonmagnetic boson mixture, quantum droplets are formed in the mean-field unstable region where the total mean-field energy is attractive.  As a result, in the Bogoliubov theory, there are imaginary excitation energies in the long wavelength limit, implying the dynamical instability.  However these excitations have little contribution to LHY energy \cite{Petrov2015}, and it is postulated that these excitations are stable after integrating out high-energy excitations.  The LHY energy is thus obtained in the Bogoliubov theory with the imaginary part neglected, and used in the construction of the extended Gross-Pitaevskii equation (EGPE).  The EGPE approach has been quite successful in interpreting experimental results \cite{Wachtler2016b,Baillie2016a,Petrov2016a,Kartashov2019}, with support  to some extent by quantum Monte Carlo calculations \cite{Saito2016a,Macia2016,Cinti2017,Bottcher2019b}. Despite the success of the EGPE, how the dynamical instability is avoided in quantum droplets, or even, whether the Bose-Einstein-condensation (BEC) state is a stable ground state has become a pressing issue.  The pairing \cite{Hu2020} and Gaussian states \cite{Wang2020} have been proposed as ground states.  Recently in a Beliaev approach,  it is found that the phonon energy is always positive after considering the interaction between spin and density fluctuations \cite{gu2020phonon}, showing that the BEC state is a stable ground state for a nonmagnetic binary quantum droplet.

In the Bogoliubov theory of a uniform dipolar Bose gas \cite{Lima2011}, in the quantum droplet region the excitation energy in the direction perpendicular to the polarization is imaginary in the long-wavelength limit, suggesting dynamic instability.  The Gaussian state was also proposed to be the ground state of dipolar quantum droplets \cite{Wang2020}. In this work, we study the excitation stability of dipolar quantum droplets by going beyond the Bogoliubov theory and calculating the excitation energy in the Beliaev formalism \cite{beliaev1958}.  In the quantum droplet region, as there are imaginary excitation energies in the Bogoliubov theory, we propose a self-consistent method to take account of the fluctuation effect from the start and avoid the complex excitation energy.  In the long-wavelength limit the excitation energy is stable with relative dipole-dipole interaction (DDI) strength $\epsilon_{\rm{dd}}<1.35$ for the $^{166}\rm{Er}$ system.  We obtain the anisotropic phonon velocity consistent with the superfluid hydrodynamic theory, but slightly different from the EGPE result due to the quantum depletion.  We propose a modified EGPE to produce the correct phonon velocity.  Finally we discuss implication of our results in experiment.

\textit{Beliaev theory of a dipolar Bose gas.} We consider a homogeneous gas of bosonic particles with mass $m$ and a finite magnetic dipole moment $d$ at zero temperature, with condensate density $n_0$. The magnetic dipoles align along the $z$ axis with the interaction potential is given by
\begin{eqnarray}
	V_{\rm{int}}(\vb{r})=g[\delta(\vb{r})+\frac{3\epsilon_{\rm{dd}}}{4\pi|\vb{r}|^3}(1-3\frac{z^2}{|\vb{r}|^2})],
\end{eqnarray}
where $g$ is the $s$-wave coupling constant, $g=4\pi\hbar^2a_s/m$, $a_{\rm{s}}$ is the $s$-wave scattering length, $\epsilon_{\rm{dd}}=a_{\rm{dd}}/a_{\rm{s}}$ is the relative interaction strength, $a_{\rm{dd}}=\mu_0d^2m/12\pi\hbar^2$ is the characteristic length of DDI, and $\mu_0$ is the magnetic permeability in vacuum. The Hamiltonian of a dipolar Bose gas is given by
\begin{eqnarray}
	\hat{H}&&=\int d^3r\hat{\psi}^{\dag}(\vb r)(-\frac{\hbar^2\nabla^2}{2m})\hat{\psi}(\vb r) \nonumber \\
	&&+\frac{1}{2}\int d^{3}r d^{3}r'\hat{\psi}^{\dag}(\vb r)\hat{\psi}^{\dag}(\vb r')V_{\rm{int}}(\vb r-\vb r')\hat{\psi}(\vb r')\hat{\psi}(\vb r),
\end{eqnarray}
where $\hat{\psi}(\vb r)$ is the boson field operator.  The Fourier transform of the interaction potential is given by
\begin{eqnarray}
	U(\vb{p})=g[1+\epsilon_{\rm{dd}}(3\rm{cos}^2\theta-1)],
\end{eqnarray}
where $\theta$ is the angle between $\vb{p}$ and the $z$ axis \cite{Lahaye2009}.

In the BEC phase, there are macroscopic number of atoms in the zero-momentum state, and the boson field operators of the zero-momentum state $\hat{a}_0$ and $\hat{a}_0^{\dag}$ can be replaced by a $c$-number $N_0^{1/2}$, where $N_0=n_0V$ is the number of bosons in the condensate and $V$ is the volume \cite{schutzhold2006}.  The Boson field operator can be separated into two parts,
\begin{gather}
	\hat{\psi}(\vb r)\equiv \hat{a}_0/V^{1/2}+\widetilde{\psi}(\vb r)\approx n_0^{1/2}+\widetilde{\psi}(\vb r), \\
	\hat{\psi}^{\dag}(\vb r)\equiv \hat{a}_0^{\dag}/V^{1/2}+\widetilde{\psi}^{\dag}(\vb r)\approx n_0^{1/2}+\widetilde{\psi}^{\dag}(\vb r),
\end{gather}
where $\widetilde{\psi}^{\dag}(\vb r)$ is the field operator of bosons outside the condensate.  In this work, we consider the dilute case where
the interaction term in the Hamiltonian can be expanded in terms of $n_0^{1/2}$.  The Bogoliubov Hamiltonian describes the quadratic fluctuation, given by
\begin{gather}
	\hat{H}_{B}=\frac{1}{2}n_0^2 VU(0)+\frac{1}{2}n_0\sum\limits_{\vb k}U(\vb k)(\hat{a}_{\vb k}^{\dag}\hat{a}_{\vb {-k}}^{\dag}+\hat{a}_{\vb k}\hat{a}_{\vb {-k}}) \notag \\
	+\sum\limits_{\vb k}[\varepsilon_{\vb k}+n_0 U(0)+n_0U(\vb k)]\hat{a}_{\vb k}^{\dag}\hat{a}_{\vb k},
\end{gather}
where $\hat{a}_{\vb k}$ is the boson annihilation operator in the momentum space.
The quasi-particle excitation energy is obtained by diagonalzing the Bogoliubov Hamiltonian,
\begin{equation}\label{BEE}
	\varepsilon_{\vb p}=\sqrt{\varepsilon^{(0)}_{\vb p}(2n_0U(\vb p)+\varepsilon^{(0)}_{\vb p})},
\end{equation}
where $\varepsilon^{(0)}_{\vb p}=\hbar^2 p^2/(2m)$ is the kinetic energy of the atom.

In the BEC phase, the boson Green's function is a matrix, defined as \cite{beliaev1958}
\begin{equation}
	{\rm G}({\vb p},t_1-t_2)=-i\langle T\{ \Psi_{\vb p}(t_1)\Psi_{\vb p}^{\dag}(t_2)\}\rangle,
\end{equation}
where $\Psi^{\dag}_{\vb p}(t)=[\hat{a}^{\dag}_{\vb p}(t),\hat{a}_{-\vb p}(t)]$, and $T\{\dots\}$ is the time-ordering operator. In the momentum and energy space, the Dyson's equation is given by
\begin{equation}
	{\rm G}(p)={\rm G}^{(0)}(p)+{\rm G}^{(0)}(p){\rm \Sigma}(p){\rm G}(p),
\end{equation}
where $p=(\vb p,p^0)$, and ${\rm \Sigma}(p)$ is the self-energy.  In the matrics of the Green's function and self-energy,
\begin{gather}
	{\rm G}(p)=\begin{bmatrix}
		G_{11}(p) & G_{12}(p) \\
		G_{21}(p) & G_{11}(-p)
	\end{bmatrix}, \\
	{\rm G}^{(0)}(p)=\begin{bmatrix}
		G_0(p) & 0 \\
		0 & G_0(-p)
	\end{bmatrix}, \\
	{\rm \Sigma}(p)=\begin{bmatrix}
		\Sigma_{11}(p) & \Sigma_{12}(p) \\
		\Sigma_{21}(p) & \Sigma_{11}(-p)
	\end{bmatrix},
\end{gather}
the subscript $11$ refers to an ingoing and an outgoing external line in the Feynman diagrams, the subscript $12$ refers to two outgoing lines, and the superscript $21$ of $\Sigma_{21}$ refers to two ingoing lines.  The free-boson Green's function is given by
\begin{equation}
	G_0(p)=1/(p^0-\varepsilon^{(0)}_{\vb p}-\mu+i0^+),
\end{equation}
where $\mu$ is the chemical potential.  From the pole-equation of the Green's function $\det|{\rm G}^{-1}(p)|=0$, the excitation spectrum can be solved.

For a dilute Bose gas, the self-energy matrix ${\rm \Sigma}({\rm p})$ can be expanded in terms of the condensate density $n_0$,
\begin{equation}
	{\rm \Sigma}({\rm p})={\rm \Sigma}^{(1)}({\rm p})+{\rm \Sigma}^{(2)}({\rm p})+\dots.,
\end{equation}
where the first-order self-energy matrix is given by
\begin{equation}\label{sigma1}
	{\rm \Sigma}^{(1)}({\rm p})=\begin{bmatrix}
		n_0U(0)+n_0U(\vb p) & n_0U(\vb p) \\
		n_0U(\vb p) & n_0U(0)+n_0U(\vb p)
	\end{bmatrix}.
\end{equation}
The chemical potential can be expanded as well, and the first-order term is given by $\mu^{(1)}=n_0 U(0)$.
The first-order Green's function is given by
\begin{eqnarray}
	G_{11}(p)&=&\frac{A_{\vb p}}{p^0-\varepsilon_{\vb p}+i0^+}-\frac{B_{\vb p}}{p^0+\varepsilon_{\vb p}-i0^+}, \\
	{G}_{12}(p)&=&-C_{\vb p}[\frac{1}{p^0-\varepsilon_{\vb p}+i0^+}-\frac{1}{p^0+\varepsilon_{\vb p}-i0^+}]  \notag \\
	&=&{G}_{21}(p),
\end{eqnarray}
where
\begin{gather}
	A_{\vb p}=[\varepsilon_{\vb p}+\varepsilon_{\vb p}^{(0)}+n_0U(\vb{p})]/2\varepsilon_{\vb p},\notag \\
	B_{\vb p}=[-\varepsilon_{\vb p}+\varepsilon_{\vb p}^{(0)}+n_0U(\vb{p})]/2\varepsilon_{\vb p},\notag \\
	C_{\vb p}=n_0U(\vb{p})/2\varepsilon_{\vb p}.
\end{gather}
From the pole of the Green's function, the quasiparticle excitation energy can be obtained,
same as that in Eq.~(\ref{BEE}) from the Bogoliubov theory.  In the droplet regime with $\epsilon_{\rm{dd}}>1$, the interaction $U(\vb p)$ is attractive in the direction perpendicular to the polarization with $\theta=\pi/2$, and the excitation energy becomes imaginary in the long wavelength limit, indicating the dynamical instability.

\begin{figure*}[] 
	\centering 
	\includegraphics[width=0.9\textwidth]{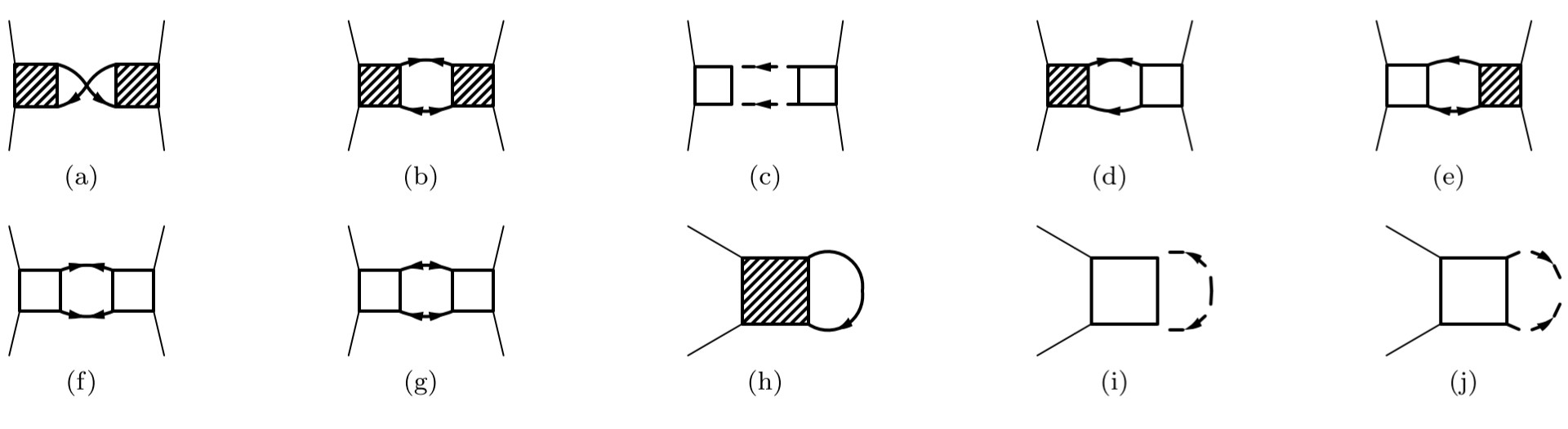} 
	\caption{Feynman diagrams of second-order self-energies. A filled rectangle denotes a sum of two rectangles, one for a direct interaction and the other for an exchange interaction. In the diagrams with four external lines, two lines are with zero momentum. The internal lines  are first-order Green's functions. The line with one arrow is the normal part of the Green's function, and that with two arrows is the anomalous part. A dashed line represents subtracting the contribution of $G^{(0)}$ to avoid double-counting.} 
	\label{sigma} 
\end{figure*}

In Ref. \cite{beliaev1958}, Beliaev calculated the second-order self-energy from one-loop diagrams and obtained the correction to excitation spectrum of a Bose gas with a $s$-wave interaction.
Here for a dipolar Bose gas, we adopt Belieav's approach to consider the fluctuation effect beyond the Bogoliubov theory, and obtain the second-order self-energy shown in Fig.~\ref{sigma}, given by

	\begin{widetext}
	\begin{gather}
		\Sigma_{11}^{(2)}(p)=\frac{1}{2}n_0\int \frac{d \vb q}{(2\pi)^3} \frac{1}{p^0-\varepsilon_{\vb k}-\varepsilon_{\vb q}+i0^+}\times\{[U(\vb p)+U(\vb k)]^2(A_{\vb k};B_{\vb q}) +2[U(\vb p)+U(\vb k)][U(\vb p)+U(\vb q)]C_{\vb k}C_{\vb q}
		\notag \\+2[U^2(\vb k)+U(\vb k)U(\vb q)]A_{\vb k}A_{\vb q}
		-2[U(\vb p)+U(\vb q)][U(\vb k)+U(\vb q)](A_{\vb k};C_{\vb q})\}\notag \\
		-\frac{1}{2}n_0\int \frac{d \vb q}{(2\pi)^3} \frac{1}{p^0+\varepsilon_{\vb k}+\varepsilon_{\vb q}-i0^+}
		\times\{[U(\vb p)+U(\vb k)]^2(A_{\vb k};B_{\vb q})+2[U(\vb p)+U(\vb k)][U(\vb p)+U(\vb q)]C_{\vb k}C_{\vb q}\notag \\
		+2[U^2(\vb k)+U(\vb k)U(\vb q)]B_{\vb k}B_{\vb q}
		-2[U(\vb p)+U(\vb q)][U(\vb k)+U(\vb q)](B_{\vb k};C_{\vb q})\}\notag \\
		+\frac{1}{2}n_0\int \frac{d \vb q}{(2\pi)^3} U^2(\vb q)(\frac{1}{\varepsilon_{\vb k}}+\frac{1}{\varepsilon_{\vb q}})
		+\int \frac{d \vb q}{(2\pi)^3}[U(0)+U(\vb k)]B_{\vb q}
		-\frac{1}{2}n_0\int \frac{d \vb q}{(2\pi)^3} U^2(\vb q)[\frac{4}{\varepsilon^{(0)}_{\vb p}-\varepsilon^{(0)}_{\vb q}-\varepsilon^{(0)}_{\vb k}+i0^+}+(\frac{1}{\varepsilon_{\vb k}}+\frac{1}{\varepsilon_{\vb q}})],\notag\\
		\Sigma_{12}^{(2)}(p)=\frac{1}{2}n_0\int \frac{d \vb q}{(2\pi)^3}(\frac{1}{p^0-\varepsilon_{\vb k}-\varepsilon_{\vb q}+i0^+}-\frac{1}{p^0+\varepsilon_{\vb k}+\varepsilon_{\vb q}-i0^+})
		\times\{[U(\vb p)+U(\vb k)][U(\vb p)+U(\vb q)](A_{\vb k};B_{\vb q}) \notag \\
		+2([U(\vb p)+U(\vb q)]^2+U(\vb q)U(\vb k)+U(\vb q)^2)C_{\vb k}C_{\vb q}
		-[U(\vb k)+U(\vb q)][U(\vb k)+U(\vb p)][(A_{\vb k}+B_{\vb k});C_{\vb q}]\} \notag \\
		+\frac{1}{2}n_0\int \frac{d \vb q}{(2\pi)^3} U^2(\vb q)(\frac{1}{\varepsilon^{(0)}_{\vb q}}-\frac{1}{\varepsilon_{\vb q}}), \label{sigma2}
	\end{gather}
where $\vb k=\vb p - \vb q$, the symbol (;) denotes a symmetrized product, $(A_{\vb k};B_{\vb q})=A_{\vb k}B_{\vb q}+B_{\vb k}A_{\vb q}$.    The second-order chemical potential satisfies the Hugenholtz-Pines theorem \cite{Hugenholtz1959}, 
	\begin{equation}
\mu^{(2)}=\Sigma_{11}^{(2)}(0)-\Sigma_{12}^{(2)}(0)=\int \frac{d \vb q}{(2\pi)^3}[U(0)+U(\vb q)]B_{\vb q}+\frac{1}{2}n_0\int \frac{d \vb q}{(2\pi)^3} U^2(\vb q)(\frac{1}{\varepsilon^{(0)}_{\vb q}}-\frac{1}{\varepsilon_{\vb q}}).
\end{equation}
The pole equation of the Green's function up to the second order is given by

	\begin{gather}
		[p^0-\frac{1}{2}(\Sigma_{11}^{+(2)}-\Sigma_{11}^{-(2)})]^2
		=[\varepsilon^{(0)}_{\vb p}+2n_0U(\vb p)+\frac{1}{2}(\Sigma_{11}^{+(2)}+\Sigma_{11}^{-(2)})-\mu^{(2)}+\Sigma_{12}^{(2)}]
		\times [\varepsilon^{(0)}_{\vb p}+\frac{1}{2}(\Sigma_{11}^{+(2)}+\Sigma_{11}^{-(2)})-\mu^{(2)}-\Sigma_{12}^{(2)}], \label{pole}
	\end{gather}
\end{widetext}
where $\Sigma_{11}^\pm(p)=\Sigma_{11}(\pm p)$.

In the long-wavelength limit, the second-order self-energy is given by
\begin{gather}
	\Sigma_{11}^{+(2)}-\Sigma_{11}^{-(2)}=2\alpha p^0, \notag \\
	\frac{1}{2}(\Sigma_{11}^{+(2)}+\Sigma_{11}^{-(2)})-\mu^{(2)}+\Sigma_{12}^{(2)}=\beta n_0U(\vb p), \notag \\
	\frac{1}{2}(\Sigma_{11}^{+(2)}+\Sigma_{11}^{-(2)})-\mu^{(2)}-\Sigma_{12}^{(2)}=\gamma \varepsilon^{(0)}_{\vb p}, \notag
\end{gather}
where $\alpha$, $\beta$, and $\gamma$ are coefficients. Since the second-order self-energies have one more power of $\sqrt{n_0}$ than the first-order self-energies, these coefficients can be treated as small quantities. To determine the leading correction to the phonon spectrum, we linearize the pole equation Eq.~(\ref{pole}), and obtain in the long-wavelength limit
\begin{equation}
	\varepsilon^{B}_{\vb p}=\sqrt{n_0U(\vb p)\varepsilon^{(0)}_{\vb p}[2+(4\alpha+2\gamma+\beta)]}.
\end{equation}

It is worth noting that each matrix element of the second-order self-energy is divergent in the long-wavelength limit, but the divergencies cancel each other in the final expression for the phonon energy in the combination $4\alpha+2\gamma+\beta$ as in the original Beliaeve theory \cite{beliaev1958}.  We separate the divergent parts, $\alpha_d, \beta_d$, and $\gamma_d$, from coefficients $\alpha, \beta$, and $\gamma$, 
\begin{gather}
	\alpha_d=\frac{n_0^2}{2}\int d\vb{q}[U(\vb{p})U^2(\vb{q})](\frac{1}{\varepsilon_{\vb{q}}^3}), \notag \\
	\beta_d=-{n_0^2}\int d\vb{q}[U(\vb{p})U^2(\vb{q})](\frac{1}{\varepsilon_{\vb{q}}^3}), \notag \\
	\gamma_d=-\frac{n_0^2}{2}\int d\vb{q} U(\vb{p})U^2(\vb{q})(\frac{1}{\varepsilon_{\vb{q}}^3}).
\end{gather}
which satisfy $4\alpha_d+2\gamma_d+\beta_d=0$.
The convergent parts of the coefficients are denoted by $\alpha_c, \beta_c$, and $\gamma_c$,
\begin{align}
	\alpha_c=&-4\sqrt{n_0 a_s^3/\pi}Q_3(\epsilon_{\rm{dd}}), \notag \\
	\beta_c=&16\sqrt{n_0 a_s^3/\pi}Q_3(\epsilon_{\rm{dd}})+\frac{32g}{U(\vb{p})}\sqrt{n_0 a_s^3/\pi}Q_5(\epsilon_{\rm{dd}}), \notag \\
	\gamma_c=&\frac{8}{3}\sqrt{n_0 a_s^3/\pi}Q_3(\epsilon_{\rm{dd}}).
\end{align}
The functions $Q_3(\epsilon_{\rm{dd}})$ and $Q_5(\epsilon_{\rm{dd}})$ describe the dipolar enhancement \cite{Bisset2016,lima2012beyond}, given by
\begin{align}
	Q_3(\epsilon_{\rm{dd}})=&\frac{(3\epsilon_{\rm{dd}})^{3/2}}{8} [(2+5y)\sqrt{1+y}+3y^2\ln\frac{1+\sqrt{1+y}}{\sqrt{y}}], \notag \\
	=&1+\frac{3}{10}\epsilon_{\rm{dd}}^2+O(\epsilon_{\rm{dd}}^3), \\
	Q_5(\epsilon_{\rm{dd}})=&\frac{(3\epsilon_{\rm{dd}})^{5/2}}{48}[(8+26y+33y^2)\sqrt{1+y}\notag \\
	& +15y^3\ln\frac{1+\sqrt{1+y}}{\sqrt{y}}], \notag \\
	=&1+\frac{3}{2}\epsilon_{\rm{dd}}^2+O(\epsilon_{\rm{dd}}^4),
\end{align}
where $y=(1-\epsilon_{\rm{dd}})/3\epsilon_{\rm{dd}}$.
The second-order chemical potential can be written explicitly in terms of these functions,
\begin{equation}
	\mu^{(2)}=\frac{40n_0g}{3}\sqrt{\frac{n_0a_s^3}{\pi}}Q_5(\epsilon_{\rm{dd}}).
\end{equation}
Thus in the Beliaev theory of a dipolar Bose gas, the phonon energy in the long-wavelength limit is given by
\begin{equation}\label{BE}
		\varepsilon^{B}_{\vb p}=\sqrt{n_0U({\vb p})\varepsilon^{(0)}_{\vb p}[2+(4\alpha_c+2\gamma_c+\beta_c)]}=v|{\vb p}|,
\end{equation}
where the phonon velocity $v$ is given by
\begin{align}\label{ps}
		v=&(\frac{n_0}{m})^{1/2}\{U(\vb p)[1+\frac{8}{3}\sqrt{n_0 a_s^3/\pi}Q_3(\epsilon_{\rm{dd}})]\notag \\
		&+16g\sqrt{n_0 a_s^3/\pi}Q_5(\epsilon_{\rm{dd}})\}^{1/2}.
	\end{align}
In comparison, in the Bogoliubov theory, the phonon velocity is given by $v_B=\sqrt{n_0 U(\vb p)/m}$.  The main difference comes from the last term in the square root in Eq. (\ref{ps}) which effectively increases the $s$-wave interaction and changes the anisotropy of the total interaction.

\textit{The modified EGPE.} The beyond-mean-field effect of a  dipolar Bose gas has also been considered in the superfluid hydrodynamic and EGPE by taking account of the LHY energy.  In the superfluid hydrodynamic theory, with the LHY energy considered, the sound velocity is given by \cite{lima2012beyond}
\begin{equation}\label{Lima}
	v_{s}=\sqrt{\frac{n}{m}\{U(\vb p)+16g\sqrt{n a_s^3/\pi}Q_5(\epsilon_{\rm{dd}})\}},
\end{equation}
where $n$ is the total boson density.  Up to the first-two orders, the total density is given by $n=n_0+n_d$, where the quantum-depletion density is given by \cite{lima2012beyond}
\begin{equation}
	n_d=\frac{8}{3}n_0\sqrt{n_0 a_s^3/\pi}Q_3(\epsilon_{\rm{dd}}).
\end{equation}
The last term inside the square root on r.h.s of Eq. (\ref{Lima}) is from LHY energy, where the density $n$ can be replaced by the condensation density $n_0$ with negligible difference of higher orders.  Thus the sound velocity given by Eq. (\ref{ps}) in the Beliaev theory is in agreement with the result in the superfluid hydrodynamic theory.

The sound velocity of a uniform dipolar Bose gas can be also obtained from the EGPE \cite{Wachtler2016b,Baillie2016a,baillie2017collective}, $$i\hbar \dot{\psi}={L}_{GP}\psi,$$
where $\psi$ is the wavefunction of the condensate, and
\begin{eqnarray}
	{L}_{GP}=-\frac{\hbar^2 \nabla^2}{2m}+\int d\vb {x'}V_{\rm{int}}(\vb{x-x'})\abs{\psi(\vb{x'})}^2\\ \notag
	+\frac{32}{3}g\sqrt{a_s^3/\pi}Q_5(\epsilon_{\rm{dd}})\abs{\psi}^3.
\end{eqnarray}
By linearizing both sides of the EGPE around the uniform solution $\psi_0=\sqrt{n_0}$, one can obtain the sound velocity given by
\begin{equation}\label{gGP}
	v_{GP}=\sqrt{\frac{n_0}{m}\{U(\vb p)+16g\sqrt{n_0 a_s^3/\pi}Q_5(\epsilon_{\rm{dd}})\}}.
\end{equation}
Compared with Eq. (\ref{Lima}), in the first term inside the square root of $v_{GP}$ is the condensation density $n_0$, not the total density $n$.  This is due to the fact that in the EGPE the mean-field energy is computed only in the condensate, whereas in superfluid hydrodynamics the mean-field energy includes also the contribution from the quantum depletion.   As this difference is of the same order of the LHY term, in principle the EGPE should be modified to take it into account.  It is negligible in the dilute system with $n \approx n_0$, but inneglectable in the system with significant quantum depletion.  Therefore to be consistent with the Belieav and superfluid hydrodynamic theories, the EGPE should be modified as  $$i\hbar\dot{\psi}={L}'_{GP}\psi,$$
where the mean-field term is modified,
\begin{widetext}
	\begin{eqnarray}
		{L}'_{GP}=-\frac{\hbar^2 \nabla^2}{2m}+\int d\vb {x'}V_{\rm{int}} (\vb{x-x'})\abs{\psi(\vb{x'})}^2[1+\frac{16}{9}\sqrt{ a_s^3/\pi}Q_3(\epsilon_{\rm{dd}})\abs{\psi(\vb{x'})}]+\frac{32}{3}g\sqrt{a_s^3/\pi}Q_5(\epsilon_{\rm{dd}})\abs{\psi}^3.\\ \notag
	\end{eqnarray}
\end{widetext}
In Ref. \cite{aybar2019temperature}, the mean-field term in the EGPE is also modified duo to quantum depletion, but the numerical prefactor is incorrect and not consistent with the Beliaev theory.

\textit{Application to quantum droplets.} In the droplet state, $\epsilon_{\rm{dd}}>1$, the total interaction $U(\vb p)$ is attractive in the direction perpendicular to the polarization and the Bogoliubov excitation energy $\varepsilon_{\vb p}=\sqrt{\varepsilon^{(0)}_{\vb p}(2n_0U(\vb p)+\varepsilon^{(0)}_{\vb p})}$ becomes imaginary in the long wavelength limit.  In comparison, in the excitation energy given by Eq. (\ref{BE}) from the Belieav theory, the $s$-wave component of the interaction is effectively enhanced.  Thus quantum fluctuation helps to stabilize the quantum droplet in region with $\epsilon_{\rm{dd}}>1$.  However in the Belieav theory, the Green's function with the Bogoliubov excitation energy is used in the computation, resulting in complex self-energy and excitation energy.  The similar situation also occurs in the EGPE approach where the LHY energy of the droplet computed from the Bogoliubov theory is complex and the imaginary part is simply ignored.  There is a problem in the logic of this method, as the Bogoliubov theory implies the dynamic instability of the uniform system and only the real part of its LHY energy is used to show the self-bound property of a finite-size system.  Here we apply a self-consistent method to take account of the effective enhancement of the $s$-wave interaction up to the leading order, and show quantum fluctuations increase the window of dynamic stability in the uniform system.

In the self-consistent approach, we assume that the quantum fluctuation effectively increases the $s$-wave coupling constant by $\delta g$ and determine it self-consistently.  The effective interaction is thus given by $$U'(\vb{p})=g'[1+\epsilon'_{\rm{dd}}(3\rm{cos}^2\theta-1)]$$ where $g'=g+\delta g$ and $g\epsilon_{\rm{dd}}=g'\epsilon'_{\rm{dd}}$.  The difference between the bare and effective interactions, $-\delta g$, serves as a counter term.  The rest computation procedures are essentially same as the Beliaev theory described in previous section except replacing $U$ by $U'$ and keeping track of the contribution from the counter term.  In this approach, the self-energy up to the second order is then given by
\begin{equation}
	{\rm \Sigma}'={\rm \Sigma}'^{(1)}+{\rm \Sigma}'^{(2)}-\delta g n_0 \begin{bmatrix}
		2 & 1 \\
		1 & 2
	\end{bmatrix},
\end{equation}
where the last term comes from the counter term, and the expression of ${\rm \Sigma}'^{(1)}$ and ${\rm \Sigma}'^{(2)}$ are the same as ${\rm \Sigma}^{(1)}$ in Eq. (\ref{sigma1}) and ${\rm \Sigma}^{(2)}$ in Eq. (\ref{sigma2}) except that $U$ is replaced by $U'$.  The chemical potential in this approach is given by
$\mu'=\mu'^{(1)}+\mu'^{(2)}-\delta g n_0$ with $\mu'^{(1)}$ and $\mu'^{(2)}$ defined in the same way as $\mu^{(1)}$ and $\mu^{(2)}$ except $U$ replaced by $U'$.  The excitation is then determined from the pole equation of the Green's function, and in the long-wavelength limit we obtain
\begin{equation}
	\varepsilon_{\vb p}^{B'}=\sqrt{n_0U'(\vb p)\varepsilon^{(0)}_{\vb p}[2+(4\alpha'_c+2\gamma'_c+\beta'_c)-\frac{2\delta g}{U'(\vb p)}]},
\end{equation}
where $\alpha'_c$, $\beta'_c$, and $\gamma'_c$ are defined similarly as $\alpha_c$, $\beta_c$, and $\gamma_c$.  The phonon velocity in the self-consistent approach is given by
		\begin{eqnarray}\label{ps'}
		v^2=\frac{n_0}{m}\{U'(\vb p)[1+\frac{8}{3}\sqrt{n_0 a_s^{'3}/\pi}Q_3(\epsilon'_{\rm{dd}})] \notag \\
		+16g'\sqrt{n_0 a_s^{'3}/\pi}Q_5(\epsilon'_{\rm{dd}})-\delta g \},
	\end{eqnarray}
where $g'=4\pi\hbar^2a'_s/m$.
We require that there is no $s$-wave enhancement to the effective interaction $U'$ which leads to the self-consistent condition
\begin{equation}
	\frac{\delta g}{g'}=16\sqrt{\frac{n_0a_s^{'3}}{\pi}}Q_5(\epsilon'_{\rm{dd}}).
\end{equation}
In the Bogoliubov theory, the dynamic instability occurs at $\epsilon_{\rm{dd}}=1$.  In the self-consistent theory, $\epsilon'_{\rm{dd}}<1$ at $\epsilon_{\rm{dd}}=1$, the system is dynamically stable.  Thus in general quantum fluctuations help to stabilize the system.

In experiments on $ ^{166}\rm{Er}$ atoms \cite{Chomaz2016}, the peak density is about $n_0=33\times10^{20}\, \mathrm{m}^{-3} $ at $\epsilon_{\rm{dd}}=1.23$.  We find that a uniform system with the same density is stable for $\epsilon_{\rm{dd}}<1.35$.  In experiments on $ ^{164}\rm{Dy}$ atoms \cite{Ferrier-Barbut2016}, $n_0=6.5\times10^{20}\, \mathrm{m}^{-3}$, the corresponding uniform system is stable with  $\epsilon_{\rm{dd}}<1.4$.   These results of the uniform case are applicable to a 3-dimensional droplet of very large size, which has not been realized experimentally so far. Preparing the droplets in a square potential well will make it easier to observe the linear phonon dispersion. In experiments \cite{Chomaz2016,Kadau2016b,Ferrier-Barbut2016,Ferrier-Barbut2016a,Schmitt2016a,Wenzel2017},  the dipolar droplets are always finite-size systems, elongated in the polarization direction, and can be formed with larger dipolar strengths, probably due to the finite-size effect which effectively increases the kinetic energy and overcomes the interaction energy in the direction perpendicular to the polarization. In the presence of a trap, the system size increases and the linear phonon dispersion is found in an EGPE study \cite{natale2019excitation}. We note that recently the sound velocity of the two-component dipolar Bose gas was derived in a hydrodynamic framework \cite{pastukhov2017}.

\textit{Concluding remarks.} In conclusion, we go beyond Bogoliubov approximation to obtain the phonon energy of a dipolar Bose gas in the Beliaev formalism and found that higher-order quantum fluctuations significantly increase the stability region of the quantum droplet.  The quantum fluctuations effectively enhance the $s$-wave interaction and change the anisotropy of the total interaction.  A modified EGPE in agreement with the Beliaev theory and superfluid hydrodynamics is proposed, which takes into account quantum fluctuation effects more completely.  We would like to thank Z.-Q. Yu and Q. Gu for helpful discussions.  This work is supported by the National Basic Research Program of China under Grant No. 2016YFA0301501.

\bibliographystyle{unsrt}
\bibliography{quantumdroplet_new}
\end{document}